\documentclass{article}

\usepackage{arxiv}
\usepackage[utf8]{inputenc}
\usepackage[T1]{fontenc}
\usepackage{booktabs} % For formal tables
\usepackage{algorithm}% 
\usepackage{algpseudocode}% 
\usepackage{hyperref}
\usepackage{amsmath}
\usepackage{xcolor}
\usepackage{xspace}
\usepackage{soul}
\newcommand{\ignore}[1]{}
\usepackage[textsize=tiny]{todonotes}
\usepackage{graphicx}
\usepackage{boldline}
\usepackage{lipsum}
\usepackage{graphicx}
\usepackage{subfig}
\usepackage{setspace}
\usepackage[export]{adjustbox}

\usepackage{adjustbox}
\usepackage{enumerate}
\usepackage{multicol}
\usepackage{multirow}
\usepackage[left]{lineno}
\usepackage{amsfonts}
\usepackage{colortbl}
\usepackage{float}
\usepackage{rotating}
\modulolinenumbers[5]
\usepackage{setspace}
\usepackage{gensymb}	% for degree symbol
\usepackage[official]{eurosym} % euro
\usepackage{siunitx}
\DeclareSIUnit{\EUR}{\text{\euro}}
\usepackage[flushleft]{threeparttable}
\usepackage{enumitem}
\usepackage{smartdiagram}
\usepackage{tabu,hhline,multirow}%% table package
\title{Comparative Analysis of Oscillating Wave Energy Converter Performance in Southern Coast of the Caspian Sea}

\author{
  Erfan Amini\\
  Coastal and Offshore Structures Engineering Group\\
  School of Civil Engineering,University of Tehran\\
  Tehran,Iran\\
  \texttt{erfan.amini@ut.ac.ir} \\
  \And
  Rojin Asadi \\
  Coastal and Offshore Structures Engineering Group\\
  School of Civil Engineering,University of Tehran\\
  Tehran,Iran \\
  \texttt{rojinasadi@ut.ac.ir} \\
  \And
  Danial Golbaz \\
  Coastal and Offshore Structures Engineering Group\\
  School of Civil Engineering,University of Tehran\\
  Tehran,Iran \\
  \texttt{Dgolbaz@ut.ac.ir} \\
  \And
  Mahdie Nasiri \\
  School of Mechanical Engineering\\
  Iran University of Science and Technology\\
  Tehran,Iran \\
  \texttt{mahdie\_nasiri@alumni.iust.ac.ir} \\
  \And
  Seyed Taghi Omid Naeeni \\
  Coastal and Offshore Structures Engineering Group\\
  School of Civil Engineering,University of Tehran\\
  Tehran,Iran \\
  \texttt{stnaeeni@ut.ac.ir} \\
  \And
  Fereidoun Amini \\
  School of Civil Engineering\\
  Iran University of Science and Technology\\
  Tehran,Iran \\
  \texttt{famini@iust.ac.ir} \\
 }
	
\begin{document}
\maketitle

\begin{abstract}

The importance of renewable energy supplies, particularly ocean wave energy, has become undeniable globally in the last two decades. Iran is not an exception due to the long barrier with deep water in the north and south of the country. Because of this potential, the first step is chosen in order to locate a suitable site for installing wave energy converters (WECs). The next step is an applicable design for converters based on conditional and geometrical parameters. In this study, locations with maximum potential are chosen, and conditional parameters are used as input data to design the oscillating converter geometrically. Next, geometric parameters are simulated by the ABAQUS program, and for verifying it, three case study sites in the Caspian sea are considered and designed in the WEC-Sim module in MATLAB to evaluate conditional and geometrical parameters simultaneously. Finally, both results are compared with each other. It is found that the Nowshahr port has more potential than Anzali and Amirabad port, and after considering the proposed design, the converter's productivity raised by 63 percent. It is also written that the flap's optimal width is recommended to be between 18 and 23 meters. Moreover, changes in damping incident forces have a direct relationship with the system's efficiency, and the value of this force is more on Nowshahr in pitch degree of freedom.
Finally, according to the software outputs, it can be concluded that the force applied to the power take-off system behind the converter flap, which is the most crucial force in evaluating the performance of the converter as the leading indicator, is derived and compared between studied sites.
    
\end{abstract}
\doublespacing
\keywords{
 Wave Energy Conversion \and  Wave Models\and  ABAQUS \and  WEC-Sim \and Renewable Energy.
}

\section{Introduction}

The limited supply of fossil fuels, the detrimental effect of such resources on the environment, and the growing entail of energy have led to a significant breakthrough in the searching process for other clean and renewable energy resources such as ocean energy. The share of these sources has increased in recent decades, which indicates the gradual and growing acceptance of this energy source from the industry. Concerning the more attention to harnessing renewable energy, especially with wave converters both in the scientific and commercial field, the world faces different converters' expending usage.
\\

Wave energy converters have various categories based on the construction site~\cite{oleinik2019site, wang2018review1-3}, the energy extraction system\cite{neshat2020hybrid,garcia2015control}, the way it is positioned against the wave \cite{henry2014two3-1,wei2016wave3-2}, or the type of motion absorbed by the system \cite{giannini2020typeofmotion,rodriguez2020typeofmotion}. In general, many studies have been done using optimization algorithms to find a suitable layout or optimum distance between converters by~\cite{neshat2018details,neshat2019new,neshat2019adaptive,neshat2020new,amini2020parametric}. Since the beginning of the twentieth century, most studies have been focused on wave energy converters, especially point absorbers such as \cite{murai2020pointabsorber,alamian2019point,aderinto2020pointabsorber,2020_performance,Neshat2020Multi}. Moreover, Some research projects surveyed oscillating WECs studied by   \cite{liu2020oscillating,erfan2019,liu2020oscillating,ruehl2020oscillators,sergiienko2020oscillating}. Henry \cite{19} evaluated the effect of different parameters affecting OWSCs performance through a laboratory study method. In his two models, seven parameters were assessed to stimulate the absorbing energy system's input power. These parameters were the wave period and the wave power, the flap's relative density, water depth, free-board of the flap, the gap between the tubes, the gap underneath the flap, and flap width. The latter parameter was one of the most critical parameters affecting the device's performance, which indicates an optimal limit to increase the device's efficiency by increasing the flap's width.
Yu \textit{et al}.  \cite{32} introduced WEC-Sim as an open-source module for modeling WECs in different wave conditions.
WEC-Sim was submitted through various platforms by validation and accurate evaluation through numerical modeling, laboratory studies, and boundary elements method for open-source hydrodynamic modeling\cite{33}. In this study, multiple parts and tools of this module are used in detail.
Renzi \textit{et al}. \cite{34} Investigated and developed the fundamental mathematical and hydrodynamic equations to describe the base-fixed rotary oscillator's movements. They also discussed previous theorems' insufficiency because the equations developed earlier were used to describe only wave energy converters' movement of the point absorber type. It was also shown that the mathematical governing equations related to the Oyster differ from the governing equations of other terminator converters.
To improve the energy absorption performance, Tom \textit{et al.} \cite{35} changed the geometry of OWSCs' cross-sectional shape, which was named a rotary oscillator converter with controlled geometry. Although this research was done by considering linear waves, the capture factor had improved. In addition to the design loads, the hydrodynamic loads were also reduced in this model. 

While in Iran, the use of renewable energy is still much lower than fossil fuels, the access to the sea in north and south of Iran has turned the country into one of the potentials of the region in using ocean energy (or marine energy). So far, various studies have been conducted on the feasibility, design, and evaluation of the construction of wave energy converter sites in Iran.
Wind data from 1986 to 1995 for 14 sites located north and south of Iran were taken by Zabihian, and Fung \cite{27} to predict the wave characteristics. According to their studies, the Chabahar site had the highest wave energy potential. This site is located on the shores of the Oman Sea, which is connected to the Indian Ocean, where the average wave energy is between 10 and 15 kW per meter. Besides, the Persian Gulf islands were known to have very high wave energy potential. Since these islands are not connected to the national grid, energy production through tidal wave energy in these islands can be economically viable.
Kamranzad \textit{et al.} \cite{28,29,29-2} surveyed 18 different sites to determine the energy distribution of waves as well as the average and maximum annual power. This study indicates that the highest average power was at the Govater station in the Oman Sea, which is due to the possibility of seasonal and tropical storms and the arrival of distant ocean waves in this region. In the Persian Gulf and the Caspian Sea, Lavan Island and Nowshahr's stations had the highest average power. To assign the most suitable converter for these possible convenient locations, Soleimani \textit{et al.} \cite{30} chose Amirabad in the Caspian sea, Asaluyeh port, and Farvar port in the Persian Gulf. Their research revealed that submerged floating elements suit the condition of the mentioned locations.
Furthermore, the Oscillating Water Column was chosen for the Farvar site on wave and shoreline conditions. Moreover, the Farvar site was the best location amongst the other two for having a maximum power output on three different converters, which were WECs connected to the seafloor. WECs included with bottom-fixed heave buoy array and WECs with bottom-fixed oscillating flap.

\begin{figure}[htb]
    \centering
    \includegraphics[width=0.85\linewidth]{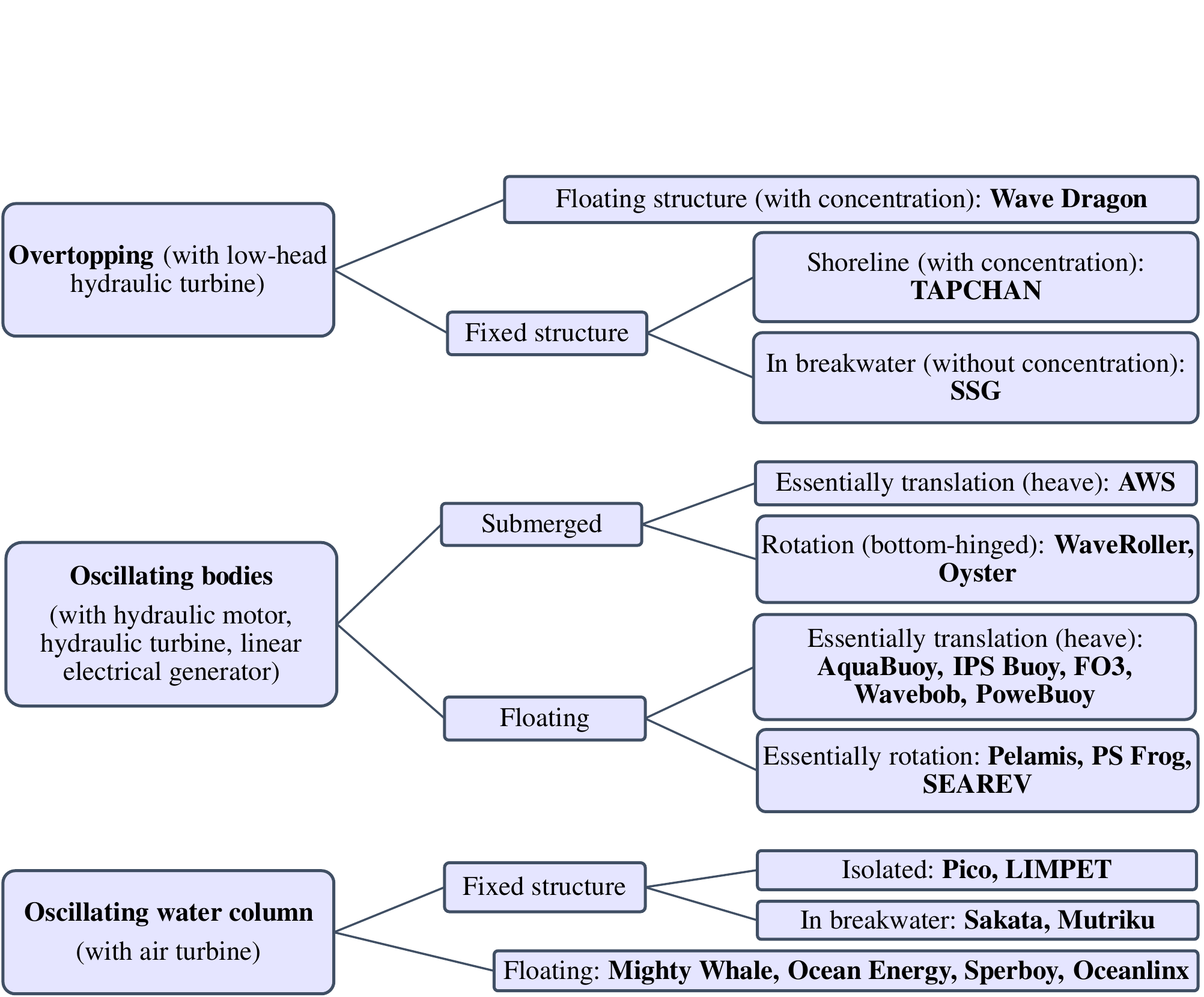}
    \caption{The various wave energy technologies.~\cite{12}}
    \label{Fig:2_4}
\end{figure}

\begin{figure}[htb]
    \centering
    \includegraphics[width=0.7\linewidth]{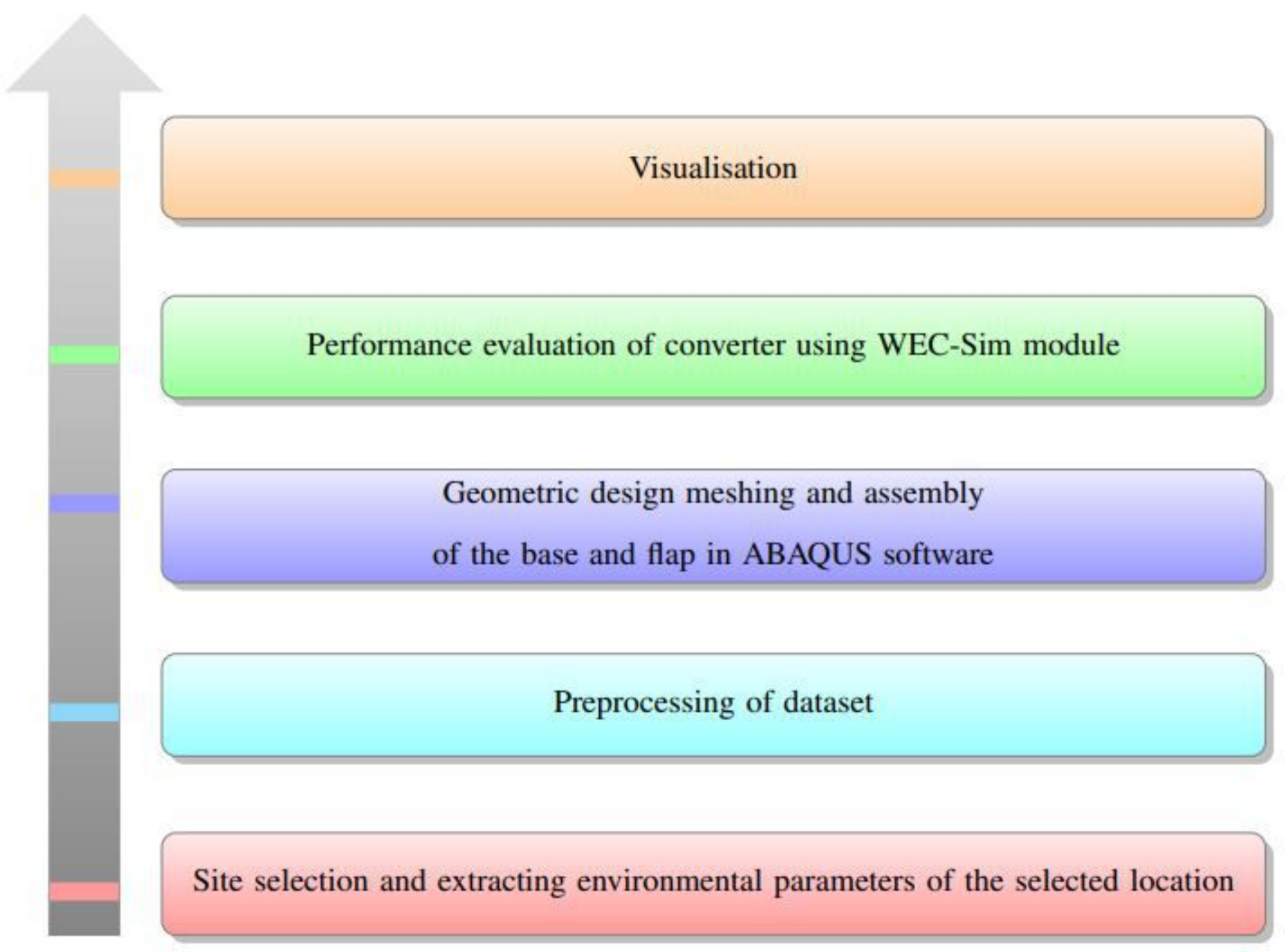}
    \caption{Steps of this research process.}
    \label{X_1_}
\end{figure}

%Research innovations 
In this study, an attempt has been made to model the converter flap's geometric design in ABAQUS software. Then, the WEC-Sim module has been used to simulate the performance of the converter. In this project, we have tried to facilitate the use of the WEC-Sim modeling module by using different toolboxes of MATLAB software such as Simulink and the solver of a multi-body dynamic; called SimMechanics\cite{62}. By considering the waves' hydrodynamic properties at each site, the feasibility of constructing an energy converter system in the three northern ports of Iran is evaluated, and a suitable design for the converter is presented. Finally, the output power performance has been investigated by a comparative study.
%wave model
One of the main decisions for exploiting energy from waves is selecting a suitable and adaptive converter to the chosen location with a high-efficiency level. In 2010, Antonio and Falcão \cite{12} presented a comprehensive classification of wave energy converters, which were divided into three categories: oscillating water columns, oscillating body systems, and over-topping converters, based on how they work. It can be said that the general technology of the developed converters while floating or fixed, can be in one of these given categories.%[12]
A comprehensive classification with regard to new technologies is presented in Figure \ref{Fig:2_4}.The given figure reveals three types of wave energy converter including over-topping \cite{aderinto2018overtoppingreview,margheritini2020overtoppingSSG,jamalabadiOvertopping}, oscillating bodies \cite{giorgi2020Oscillatingbodiesreview,penalba2017Oscillatingbodies,li2020Oscillatingbodies}, and oscillating water column \cite{falcao2016OWCreview,falcao2014OWC,bailey2016waveOWC_F}. Each of them is divided into two subsets, such as Fixed structure, Floating, or Submerged ones.
\\
This paper attempts to design the energy converter's geometric parameters according to the environmental parameters obtained for three ports (Nowshahr, Amirabad, and Anzali) by creating a geometric model of the energy converter for each of the three areas in ABAQUS software. Finally, we evaluate the oscillating surge wave energy converter's performance using geometric parameters and environmental parameters in WEC-Sim software. Figure \ref{X_1_} describes the process of our study.

This paper is organized as follows. Section \ref{Sec-rm} represents the research method, which is divided into six subsections. As mentioned earlier, this paper stands on searching for a suitable site for installing WECs in the north of Iran, so the study area is discussed in subsection \ref{SubSec-datac}. All the details about the wave energy converter used in this model are introduced in subsection \ref{SubSec-WaveEnergyModel}. The governing equations will be explained in subsection \ref{SubSec-goveq}, following a description of the Boundary Element Method (BEM) in subsection \ref{SubSec-BEM}. Setting up the simulation of this research, which is by WEC-Sim module described in subsection \ref{SubSec-PerformanceAssessmentCriteria}, follows the converter's details and optimal dimensions in subsection \ref{SubSec-geometric}. Section \ref{Sec-result} presents the output of the study properly for all three sites. Finally, in the last section, section \ref{Sec-conclusions}, some conclusions are drawn.

\section{Research method}
\label{Sec-rm}
In this section, we attempted to discuss collecting the data and performing the model's initial requirement. Furthermore, when data is processed, governing equations were implied as numerical modeling. Furthermore, the boundary element method provides the initial demands of the WEC-Sim module. Finally, a description of the geometric design provided as follows.

\subsection{Data Collection}
\label{SubSec-datac}
\begin{figure}[htb]
    \centering
    \includegraphics[width=0.8\linewidth]{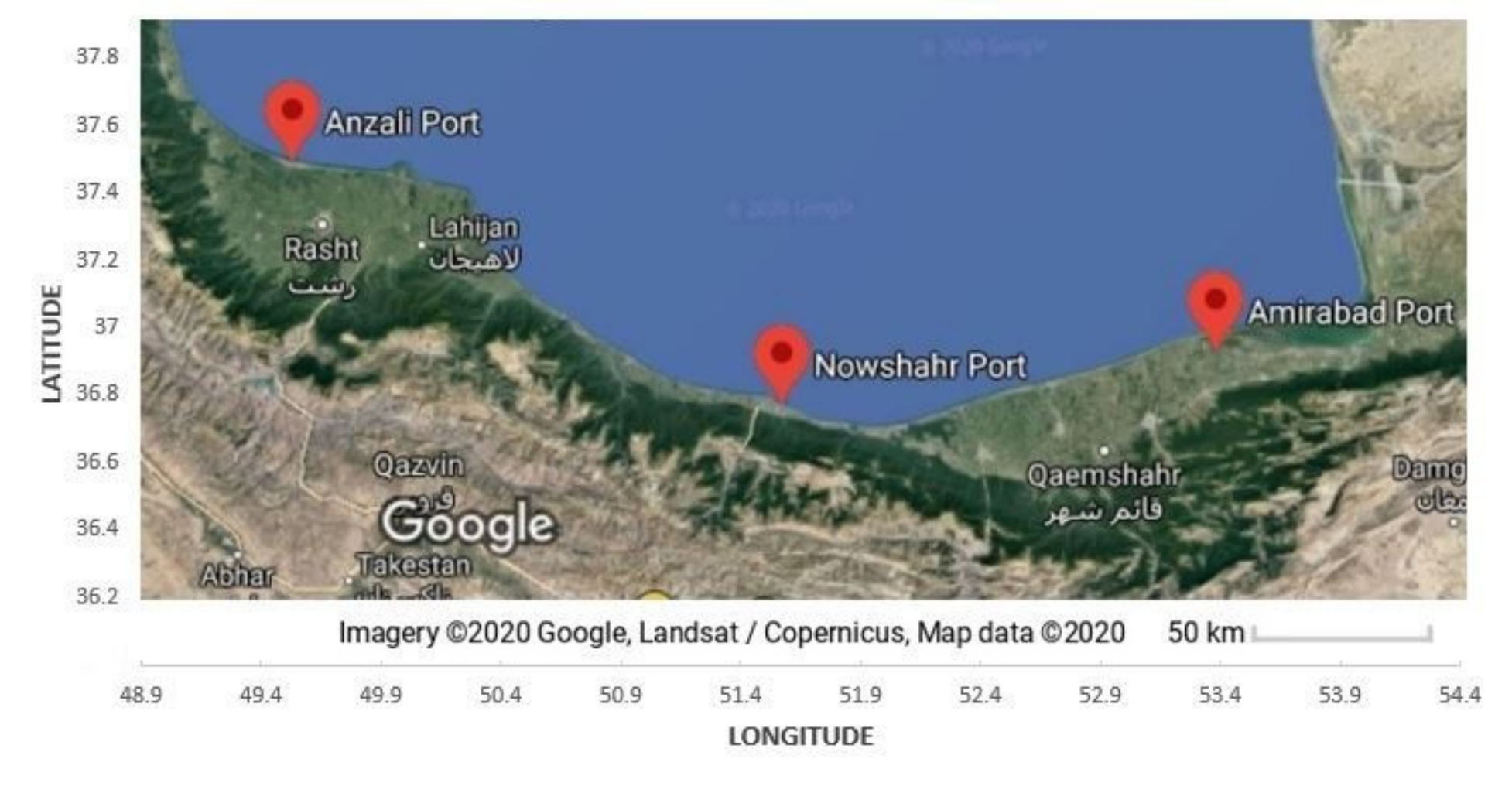}
    \caption{The locations of three ports in the south of the Caspian Sea.~\cite{21}}
    \label{3_13}
\end{figure}

Based on the research of Kamranzad \textit{et al.}, it can be concluded that Nowshahr port (in the southern part of the Caspian sea) has the potential to install WECs \cite{28}\cite{29}\cite{29-2}. Thus Nowshahr port is chosen for the case study of this research. Also, Anzali port in the southwest of Caspian sea and Amirabad port in the southeast of Caspian sea are chosen to compare results based on their wave energy capacity \cite{amini2019investigating}. The location of these three ports is shown in figure \ref{3_13}.
An extensive set of wave data is needed for the sites of research. Therefore the unprocessed metadata from SWAN modeling in the Caspian Sea area in the authority of the Atlas project of Iranian waves is given from the Iranian National Institute of Oceanography and Atmospheric Science. This dataset included Bathymetric data of the site ($h$), a period of waves ($T_i$), wave height ($H_i$), and direction of wave ($\alpha$) in time series of 27 years period from 1980 to 2007 \cite{golshani}. The given time-series data for each point was per hour, which can be seen in all the model's 2700 points. It is noted that almost 89000 data each year were used, which in total would be near 2400000. 
A code is written in MATLAB to preprocess the data to analyze and process this dataset. Then after giving the mentioned data within the area, it returns a related dataset. Next, data is classified in terms of location. It is saved in a separate file to undergo the prepossessing statistical method to eliminate the unreliable data. Moreover, the annual average of the dataset is calculated for aggregating time series \cite{amini2019investigating}. In this study, data was chosen in sectors with a radius of 0.2 degrees around the center of each port, which the table \ref{tab:Long-Lat} reveals the details. 
%page 83: table
\begin{table}[]
    \centering
\scalebox{0.7}{
}
\begin{tabular}{ccccc}
 \toprule
\textbf{Port's Name}& \textbf{Minimum Longitude}& \textbf{Maximum longitude}& \textbf{Minimum Latitude}& \textbf{Maximum Latitude}
\\ \hline	
\textbf{Anzali}&  49.26 & 49.66& 37.44 & 37.84 \\ 
\hline
\textbf{Nowshahr} & 36.6 &37.00 &51.30 &51.70 \\ \hline
\textbf{Amirabad}& 53.17& 53.57 & 36.81 & 37.21 \\ 
\bottomrule
	\end{tabular}
    \caption{Range of Longitude and Latitude for each selected site}
    \label{tab:Long-Lat}
\end{table}

After extracting the data from three regions, the wave height data was converted to the significant wave height with the following equations. Finally, their annual average for each region was calculated.

After extracting the data from three regions, wave height data ($H_i$) must be transformed into significant wave height ($H_s$) with the following equations \cite{60}. Then their annual average is calculated for each location. 
\begin{equation}
H_{rms}=\sqrt{\frac{1}{N}{\sum_{i=1}^n H_i^2}} 
\end{equation}

\begin{equation}
H_s= \sqrt{2}H_{rms}
\end{equation}

\subsection{Wave Energy Model}
\label{SubSec-WaveEnergyModel}
To absorb the waves' energy, a device is needed to absorb the maximum energy contained in them when the waves collide in the right direction. This device is called a wave energy converter and consists of several parts. On the other hand, waves have different behavioral characteristics. Therefore, the converter should be designed and implemented according to the type of shore and the waves' hydrodynamic characteristics, and the energy converter's operating limitations. 

In general, wave energy converters are divided into three categories: attenuators, point absorbers, and terminators according to their dimensions (compared to the wavelength of the collision) as well as their type and location (relative to the direction of wave propagation) \cite{16}. Converters that are long and parallel to the direction of wave propagation are known as attenuators. Converters perpendicular to wave propagation, with equal to or larger dimensions than the wavelength, are included in the terminators' category. Also, adsorbents, whose dimensions are less than one wavelength, are known as point absorbers \cite{12}.

Several factors for choosing the type of the wave energy converter are scrutinized, including the prospect of research \cite{11}, maintenance and installation cost \cite{50}, and the optimal degree of freedom \cite{51}. The final decision is made by considering previous factors. Due to the lack of accessing enough information, the importance of these factors is assumed to be equal. In sum, considering all of the mention factors, Oscillating Surge WEC is the better converter. Thus, our decision is made based on the discussed results.

\subsection{Governing Equations}
\label{SubSec-goveq}
The WEC-Sim software uses Cummins' equations in the modeling process to calculate the motion response function of the oscillator. In 1962, Cummins presented equations for the motion response of floats and ships against waves at 6 degrees of freedom \cite{65}. 
The main Cummins' equation is as follows:

\begin{equation} \label{26}
m\ddot{X}=F_{ext}+F_{rad}+F_{PTO}+F_{v}+F_{m}
\end{equation}

Where $\ddot{X}$ is the velocity vector (either displacement or rotation) of the device, $m$ is the mass matrix, $F_{ext}$ is the wave excitation force vector, $F_{rad}$ is the wave radiation force vector, $F_{PTO}$ is the power take-off force vector, $F_{v}$ is the viscous damping force vector, and $F_{m}$ is the resulting force vector from mooring system. 
$F_{ext}$ and $F_{rad}$ are calculated using the values provided by the boundary element method (BEM) solver in the frequency range. The radiation force consists of an added mass and a wave attenuation associated with the acceleration and velocity of the floating body, respectively. Suppose the floating body is constrained in the direction of wave propagation. In that case, the hydrodynamic force of the waves affecting the body is called the \textit{wave excitation force}, which in three directions of freedom will include three forces and three wave excitation torques \cite{66}.

WEC-Sim can be used for regular and irregular wave simulations, but $F_{ext}$ and $F_{rad}$ are calculated separately for the sinusoidal steady-state response scenario and stochastic wave simulations. The sinusoidal steady-state response is often used for simple wave energy converter designs with regular input waves. For stochastic wave simulations or any other simulation in which the effects of fluid time history are required, the use of the integral convolution method on a floating mass is recommended \cite{15}.

The convolution integral calculation based on Cummins' equation is used to include the effect of fluid time history on the system. Therefore, the radiation force can be calculated as follows \cite{62}.

\begin{equation}
F_{rad}=-A_{\infty}\ddot{X}-\!\!\int_{0}^{t}K(t-\tau)\ddot{X}_{\tau}\hspace{0.15cm}d\tau
\end{equation}

Where $A$ is the mass-added matrix at the infinite frequency, and $K$ is the impulse response function.\\
In irregular waves, the free surface height is made up of several regular wave components linear superposition, often determined using a wave spectrum. This spectrum describes the wave energy distribution over a wide range of wave frequencies, characterized by wave height $(H_{s})$ and a peak period $(T_{P})$. Irregular excitation force can be calculated as follows \cite{15}:
\begin{equation}
F_{ext}=\Re[R_{f}\int_{0}^{\infty}\!\!\!\sqrt{2S(\omega_{r})}F_{x}e^{i(\omega_{r}t+\phi)}\hspace{0.15cm}d\omega_{r}]=\int_{-\infty}^{+\infty}\eta_{\tau}f_{e}(t-\tau)\hspace{0.15cm}d_{(\tau)}
\end{equation}
\\
Where $\Re$ denotes the real part of the equation, $R_f$ is the ramp function, $F_x$ is the excitation vector consists of amplitude and phase of the wave, $S$  is the wave spectrum, $\phi$  is the stochastic phase angle, $\eta_{\tau}$  represents water elevation and $f_e$  is the element of the force vector.
A ramp function $(R_{f})$ is necessary to avoid initial transient oscillations in the short initial simulation times and calculate the wave's excitation force. 
if $t$ is considered as the simulation time and $t_{r}$ is the initial oscillation adjustment time, it is better to consider the value of $t_{r}$ equal to $25T_{P}$ for a more stable numerical solution. According to $T_{P}$ values in the various ports under study (which are about 4 seconds), the amount of ramp time in this study is considered equal to 100 seconds \cite{32}.
\\
The absorbed power in the power take-off mechanism is represented as a mass damping system as follows \cite{65}:
\\
\begin{equation}
P_{PTO}=(K_{PTO}X_{rel}\dot{X}_{rel}+C_{PTO}\dot{X}_{rel}^2)=C_{PTO}\dot{X}_{rel}^2
\end{equation}

In which $K_{PTO}$ is PTO stiffness, $C_{PTO}$ is PTO damping, $X_{rel}$ and $\dot{X}_{rel}$ are relative displacement and velocities. It should be mentioned that the relative motion and velocity have a $\dfrac{\pi}{2}$ phase difference, That is the reason for eliminating the stiffness sentence in the equation.

\subsection{Geometric and Numerical Design}
\label{SubSec-geometric}
%SIMULATION SET-UP
Up to this moment, four programs have been used to simulate WECs called WEC-Sim, ANSYS-AQWA, OpenFOAM, and FLOW-3D. For simulating currents around point absorber converters, ANSYS-AQWA is the most probable choice; however, the other three are recognized for simulating oscillating WECs in recent literature. In FLOW-3D, if a converter in the flume has been used, water elevation will fall gradually over time, making output data errors in the long term. Furthermore, In the OpenFOAM, the mesh motion for fluid around the flap does not match precisely with real motions; therefore, a cylinder of moving mesh is entailed, and it needs to implement further codes\cite{56}. Moreover, this program cannot read files with \textit{.stl} format, which is the format of much common design programs\cite{36}.
Given these mentioned deficiencies, WEC-Sim seemed to be a more useful program for this study.

WEC-Sim, which is an open-source wave energy converter simulation tool, is financed by the U.S Department of Energy and developed by NREL (National Renewable Energy Laboratory) and SNL (Sandia National Laboratories). This code is implemented in MATLAB and Simulink using SimMechanics, which is the solver of a multi-body dynamic problem \cite{62}.
WEC-Sim can simulate rigid bodies, power take-off (PTO) systems, and mooring systems. Hydrodynamic forces are modeled based on the diffraction and radiation method. Simultaneously, the dynamic of a time-dependant system is considered by solving motion governing equations in every six degrees of freedom on WEC. The diffraction and radiation method generally computes hydrodynamic forces from a boundary element method (BEM) solver in frequency domain \cite{New01}.
\\
According to the project, necessary changes are required for using BEM codes, input variables, and output. Next, this module was installed due to access to the primary source code, added to Simulink's library. By considering the recommendation of recent literature as well as the technical reports of Oyster in Scotland in 2007, the height of the converter's flap, after determining the depth of the optimal points, will be equal to the depth plus one-meter free-board\cite{62,63,19}. Optimal thickness is presumed 1.8 meters in the designing process, and the oscillator's width is determined from Folley, and Wittaker's graph \cite{64}. 
\begin{figure}[htb]
    \centering
    \includegraphics[width=0.5\linewidth]{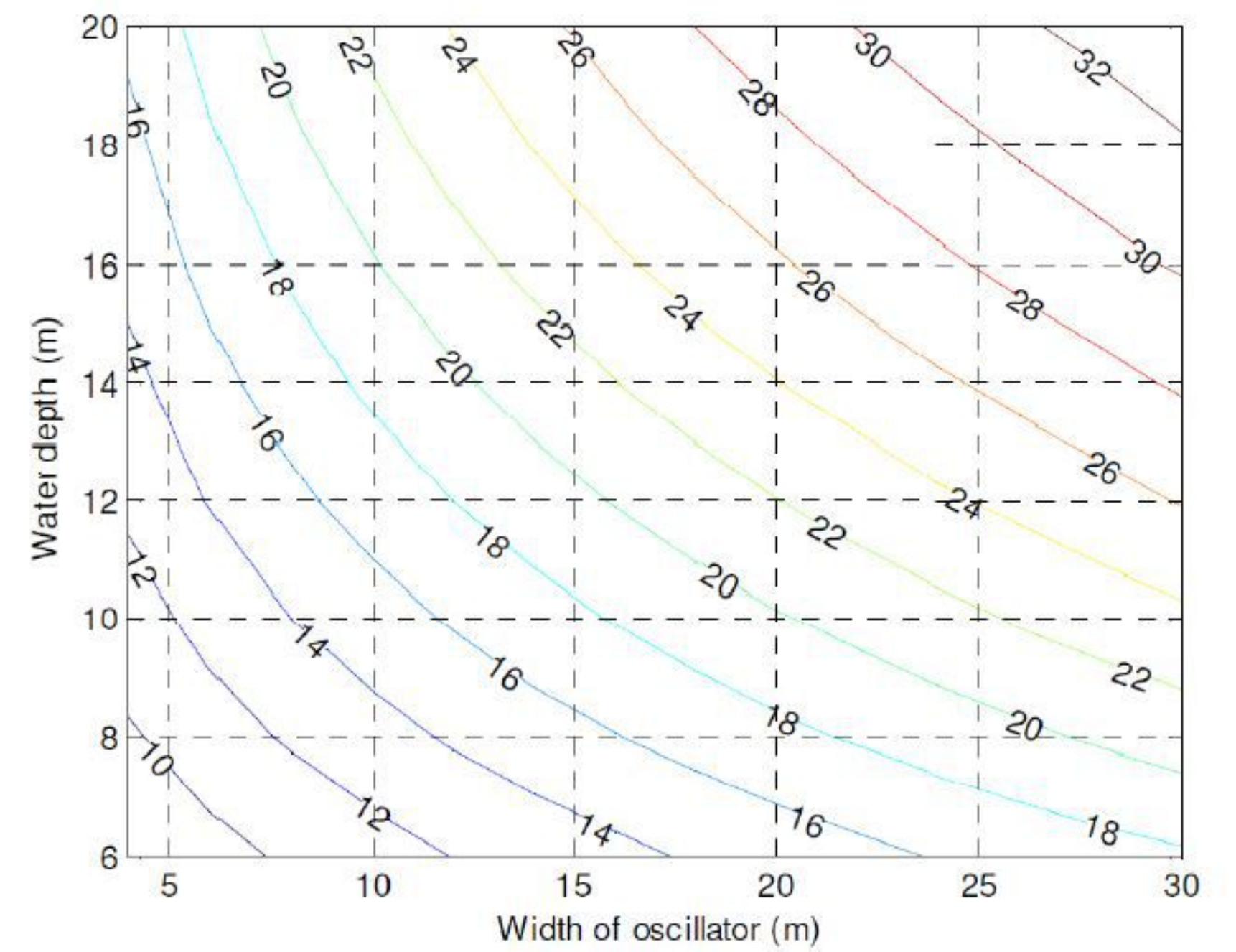}
    \caption{Determining the width of the oscillator based on the free swing period and water depth. ~\cite{64}}
    \label{3_19}
\end{figure}
It is figured out from the plot that two factors are needed to find the oscillator's width: water depth and free swing period. 
The first one can be found in the location's parameters, and the latter can be computed from \eqref{T_N} equation based on radiation relations.
 \begin{equation} \label{T_N}
 T_N= 2\pi \sqrt{I+K_p}
 \end{equation}
Where $K_p$ is rotational stiffness induce from the PTO system and $I$ is the rotational moment of inertia.

\begin{figure}[htb]
    \centering
    \includegraphics[width=0.6\linewidth]{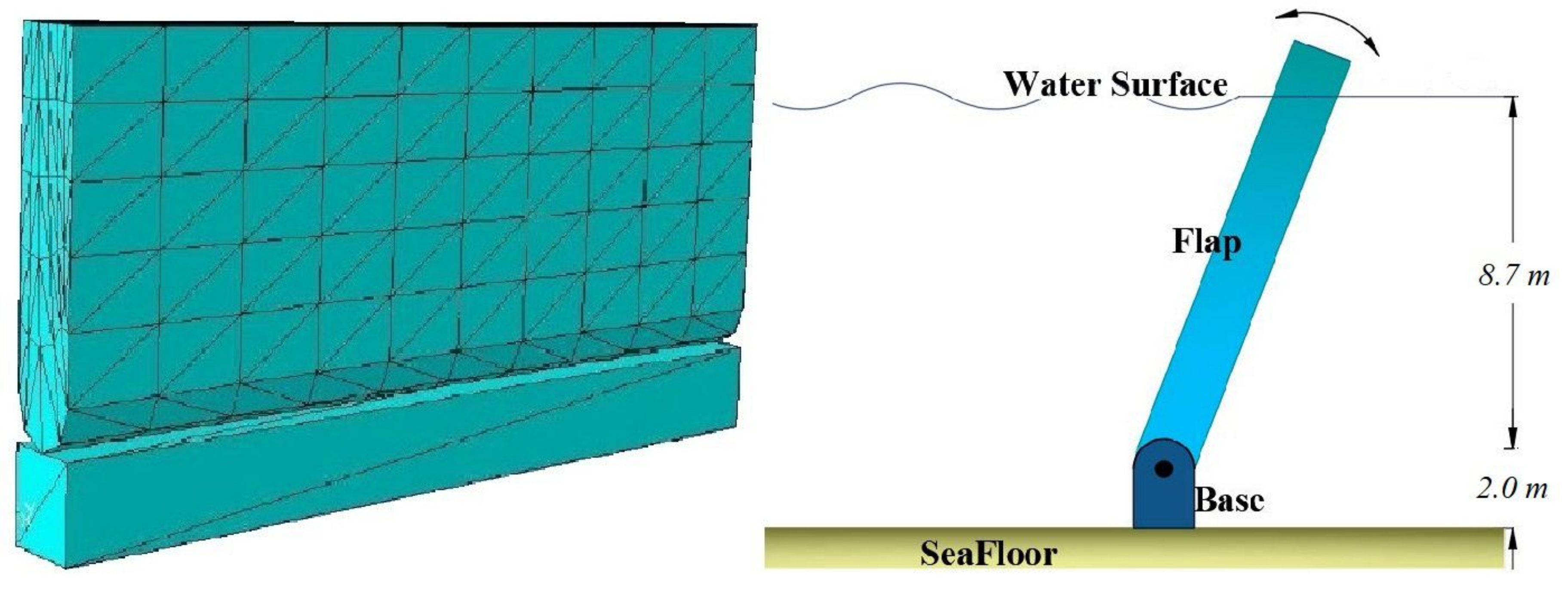}
    \caption{The designed geometrical shape of OSWEC for Nowshahr port. ~\cite{62}}
    \label{3_18_1}
\end{figure}

To introduce the converter's operation types to WEC-Sim software, specifications of the converter, types of its connectors, motion restriction, bodies' degrees of freedom, and PTO system should be given to Simulink's software module. There are four steps based on the software's manual recommendation, and the specification of motion in the rotational oscillator is described to the Simulink module:
(i) Two rigid body blocks for identification of the  oscillator's flap and base
(ii) A global reference block 
(iii) A fixed constrained block in every six degrees of freedom for the oscillator's base
(iv) A rotational PTO block for the converter's flap.
Finally, after supplying blocks' connection, the final Simulink model for oscillating surge WEC is completed.
It is needed to choose the flap's width and height according to the relations and environmental(sea-state) parameters to find the optimal dimensions of the oscillator's flap. Table \ref{tab:Table 4-2} introduces mentioned parameters.
\begin{table}[htb]
	\small
	\caption{Designed Geometrical parameters of the Oscillator's Base and Flap}
	\label{tab:Table 4-2}
	\begin{center}
		\begin{tabular}{|c|c|c|}
			\hline
			\multirow{2}{*}{\textbf{Dimension's Name}} &  \multicolumn{2}{c|}{\textbf{Amount} }\\
			\cline{2-3}
			  & \textbf{Width} & \textbf{Height}\\ \hline
			%Width of the Oscillator's Flap ($m$) &  \multicolumn{2}{c|}{ depends on the port's hydrodynamic parameters}  \\ 
			%Height of the Oscillator's flap ($m$) from rotation point &  \multicolumn{2}{c|}{ depends on the port's hydrodynamic parameters}  \\ \hline
			Optimal geometry of the Oscillator's flap in Nowshahr port &  21 & 7.2  \\ \hline
			Optimal geometry of the Oscillator's flap in Anzali port & 18 & 8.7  \\ \hline
			Optimal geometry of the Oscillator's flap in Amirabad port & 23 & 7 \\
			\hline
			Thickness the of Oscillator's Flap ($m$)&  \multicolumn{2}{c|}{1.8} \\ \hline 
			Width of the Oscillator's Base ($m$)&  \multicolumn{2}{c|}{18} \\ 
			\hline 
			Thickness of the Oscillator's Base ($m$) &  \multicolumn{2}{c|}{1.8} \\ \hline 
			Height of the Oscillator's Base ($m$) &  \multicolumn{2}{c|}{1.8} \\ \hline 
			The distance between center of rotation and seafloor**** ($m$)&  \multicolumn{2}{c|}{2} \\ \hline 
			Moment of Rotational Inertia (Pitch) ($kg.m^2$) &  \multicolumn{2}{c|}{1850000} \\ \hline 
			Damping coefficient of PTO system ($\frac{Nsm}{rad}$) &  \multicolumn{2}{c|}{12000} \\ \hline 
			Mass ($kg$) &  \multicolumn{2}{c|}{127000} \\ \hline 
			Distance between center of mass and surface of the flap ($m$) &  \multicolumn{2}{c|}{ -3.9} \\ \hline
			
		\end{tabular}
	\end{center}
\end{table}
 %step 7
 For simulating the Oscillating wave surge converter in WEC-Sim, firstly, it needs to design the flap and base of the oscillator in CAD programs. These programs' output is taken in two separate files in \textit{.stl} format as a geometrical input for WEC-Sim. ABAQUS software has been chosen to design the geometrical parameters of the converter. It is worth noting that the dimensions of the designed model in ABAQUS can be seen in figure \ref{3_18_1}.
When the described procedure is done, the measures of width and height of the flap are obtained in the three locations of research.

Next, the base and flap were designed in ABAQUS CAE 2019 software, released by DS SIMULIA. Due to the specification of Oyster converters installed in Orkney shores of Scotland, the material of flap is chosen to be Aluminium with a density of 2700 kilogram per cubic meter with uniform mass distribution. Other parameters such as elastic modulus and Poisson's ratio are 69 gigapascals (GPa) and is 0.33, respectively. When the flap and base design are completed in CATPart form, these two models are assembled. And to generate mesh on them, surface meshing with triangular elements is chosen regarding the BEM solver. The relative results are revealed in section \ref{Sec-result}.

\subsection{Boundary Element Method}
\label{SubSec-BEM}
In this method, the governing differential equations are converted into integral unions applied to the surface or boundary. These integrals are numerically integrated on the boundary. In this approach, the boundary is divided into small parts (boundary elements). 
Also, because all approximations are surface-specific, the boundary element method can model areas involving drastic changes of variables with better accuracy than the finite element method \cite{68}.
We have used this method to perform the necessary hydrodynamic data calculations before starting the WEC-Sim operation. The software works first to solve the boundary element method through the open-source code, NEMOH.

This code applies fluid environmental conditions and meshes the converter flap and base by receiving the converter geometry file. Then, it calculates the hydrodynamic coefficients in the frequency domain, which include the velocity potential and the pressure field, using the boundary integral equation method (also called the panel method). Next, the values of the added mass matrix $(A_\infty)$, the impact response function $(K)$, and the wave excitation vector $(F_x)$ are obtained \cite{15}. 
 
The functional steps of the boundary element method codes are: Receiving NEMOH outputs, Calculation of Impulse response functions (IRFs) for wave radiation and excitation, Calculation of state-space realization coefficients (calibration) for wave radiation response functions, and finally, Data storage in the form of hierarchical data format(HDF5) \cite{15}.

\subsection{Performance Assessment Criteria}
\label{SubSec-PerformanceAssessmentCriteria}
\begin{figure}[htb]
    \centering
    \includegraphics[width=0.85\linewidth]{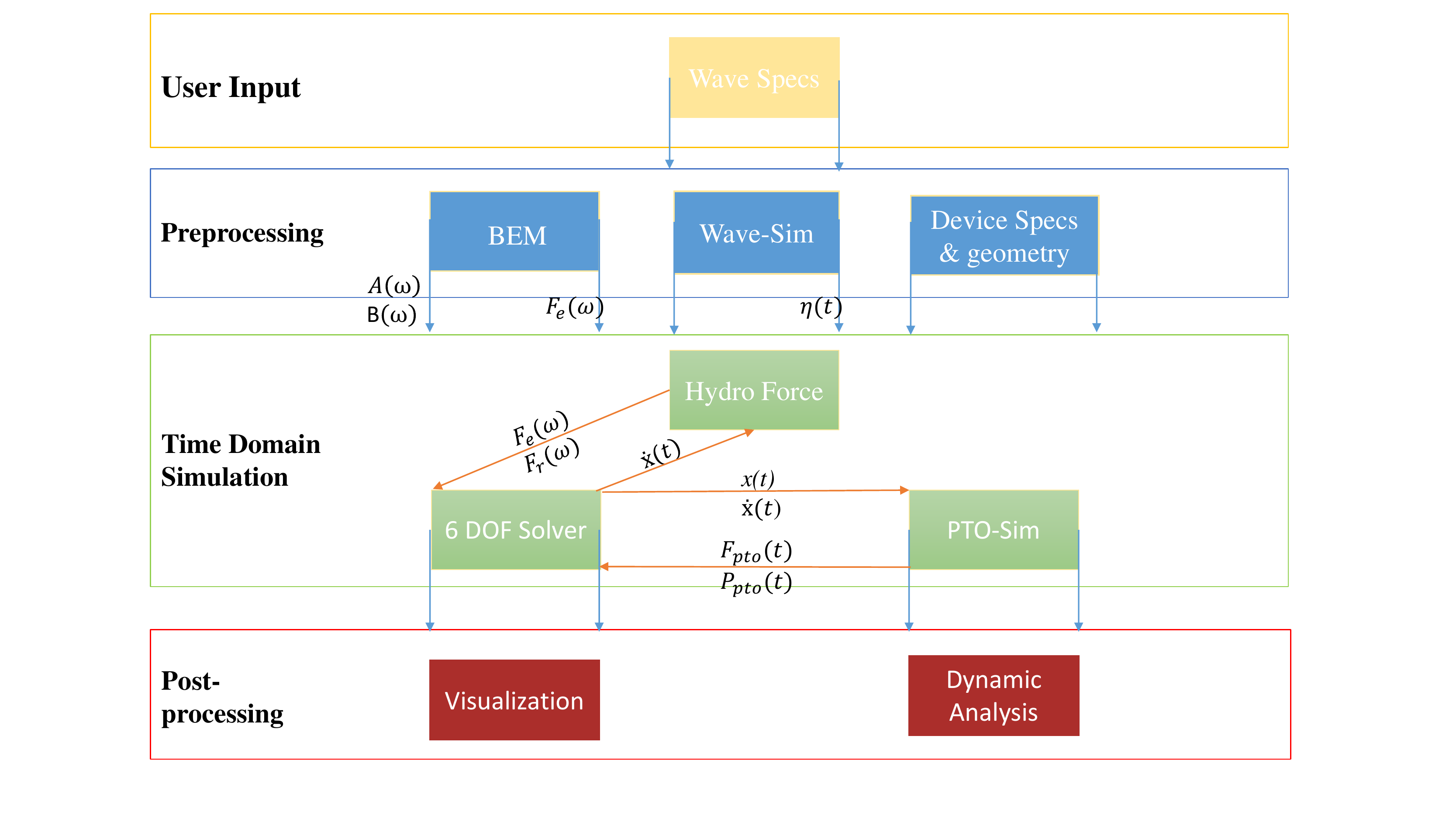}
    \caption{Numerical simulating in WEC-Sim module.(Adapted from\cite{32})}
    \label{3_211}
\end{figure}

According to the given explanations, the general modeling process by WEC-Sim software is shown in Figure \ref{3_211}.

As shown in figure \ref{3_211}, the converter's specifications and geometry and environmental conditions (in the form of wave spectrum) are given as the software's input. The software calculates the structure's hydrodynamic parameters in the pre-processing section and models the waves according to the input parameters. In the main processing section, the structure's response characteristics to the collision of waves in the time domain are calculated. Finally, in the post-processing part, the required diagrams and analyzes are extracted. The following equations are also used to calculate the efficiency of converters \cite{19}:

\begin{equation} \label{p_i}
P_i=\dfrac{\mathcal{F}}{C_{g}t}
\end{equation}

In equation \eqref{p_i}, $P_i$ is the longitudinal power of the incident waves, $\mathcal{F}$ is the wave energy flux, and $C_g$ is the wave group velocity.
\begin{equation}
\mathcal{F}=\dfrac{\rho g \omega_{i} A_{i}^2}{4k_{i}} (1+\dfrac{2k_{i}d_{i}}{\sin h (2k_{i}d_{i})})\sin\alpha_i
\end{equation}

\begin{equation}
C_g=\dfrac{\omega}{k_{0}} . \dfrac{1}{2} [1+\dfrac{2k_{0}h}{\sinh (2k_{0}h)}]
\end{equation}

 Also, $k_0$ is wave number, $h$ is water depth, and $H$ is the height of incident waves. 

\begin{equation} \label{c_f}
C_f=\dfrac{P_{PTO}}{P_i}
\end{equation}

In equation \eqref{c_f} the capture factor or total efficiency per unit length $(C_f)$ indicates the length of the wave crest, all of which is extracted by the device. This parameter is obtained by dividing the absorption power by the power of the impact waves, and its unit is in $m^{-1}$\cite{19}. 

\section{Result and Discussion} \label{Sec-result}
 
%118-124:
 \subsection{Interaction Implied Forces}
\begin{figure}[htb]
	\centering
	\subfloat [Surge]{
    	\includegraphics[height = 5cm, width = 7.5cm]{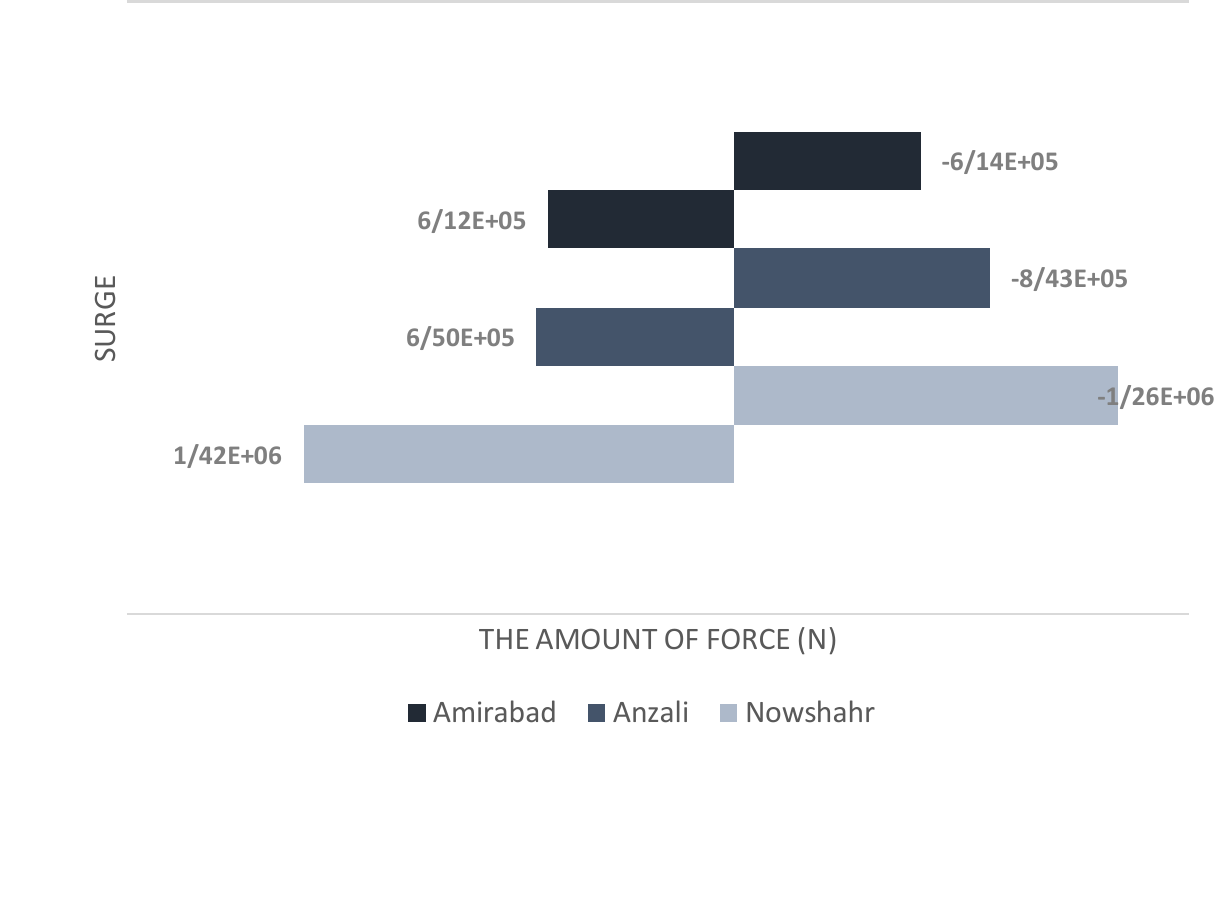}
      		} 
      		\hfill
  % \hspace{0.5cm}
	\subfloat[Pitch]{
   		\includegraphics[height = 5cm, width = 7.5cm]{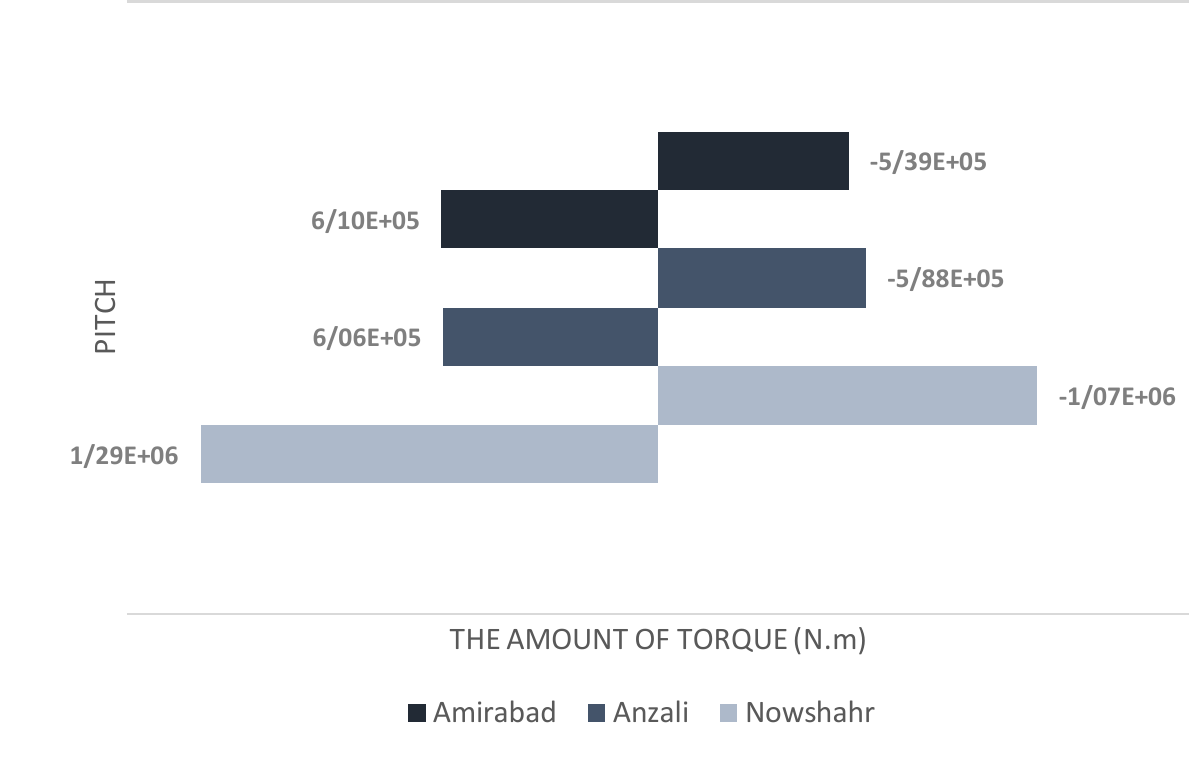}
 	  	}
 	  	\caption{Comparison of the wave power of radiation damping force both in surge and pitch in all locations. (from up to down: Amirabad, Anzali, and Nowshahr)}
	\label{4_17}
\end{figure}

 In this part, all encountered forces and torques are obtained from the WEC-Sim module. Based on equation\eqref{26}, the considered forces are excitation force, radiation resistance (or wave damping), PTO forces, and added mass force. Each of these forces is calculated both in Surge and Pitch degrees of freedom. It is worth to note that Viscous, Buoyancy forces, and mooring forces are not modeled because of their negligible effect on the converter's displacement. In studying the radiation damping force of a wave, this force will have different values for the two modes of moving the converter flap backward (collision of a wave crest) and moving the converter flap forward (collision of a wave through).
For each port in two degrees of freedom of Pitch and Surge, software outputs are obtained, and the values are presented in Figure \ref{4_17}. As this shows, firstly, in all three ports, the radiation damping force values are larger than the radiation damping torque. Since the converters' rotational speed values are less than their linear velocities \cite{69}, this seems logical.

\begin{figure}[htb]
	\centering
	\subfloat [Surge]{
    	\includegraphics[height = 5cm, width = 7.5cm]{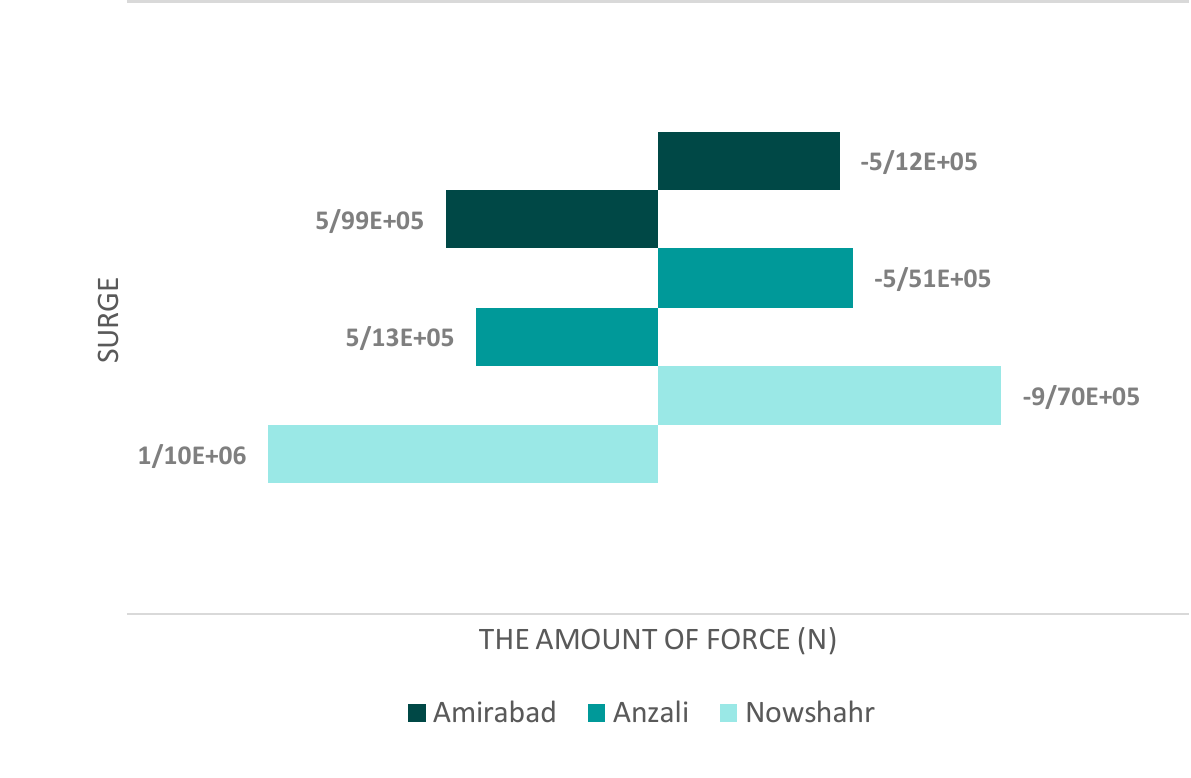}
      		} 
  % \hspace{0.5cm}
  \hfill
	\subfloat[Pitch]{
   		\includegraphics[height = 5cm, width = 7.5cm]{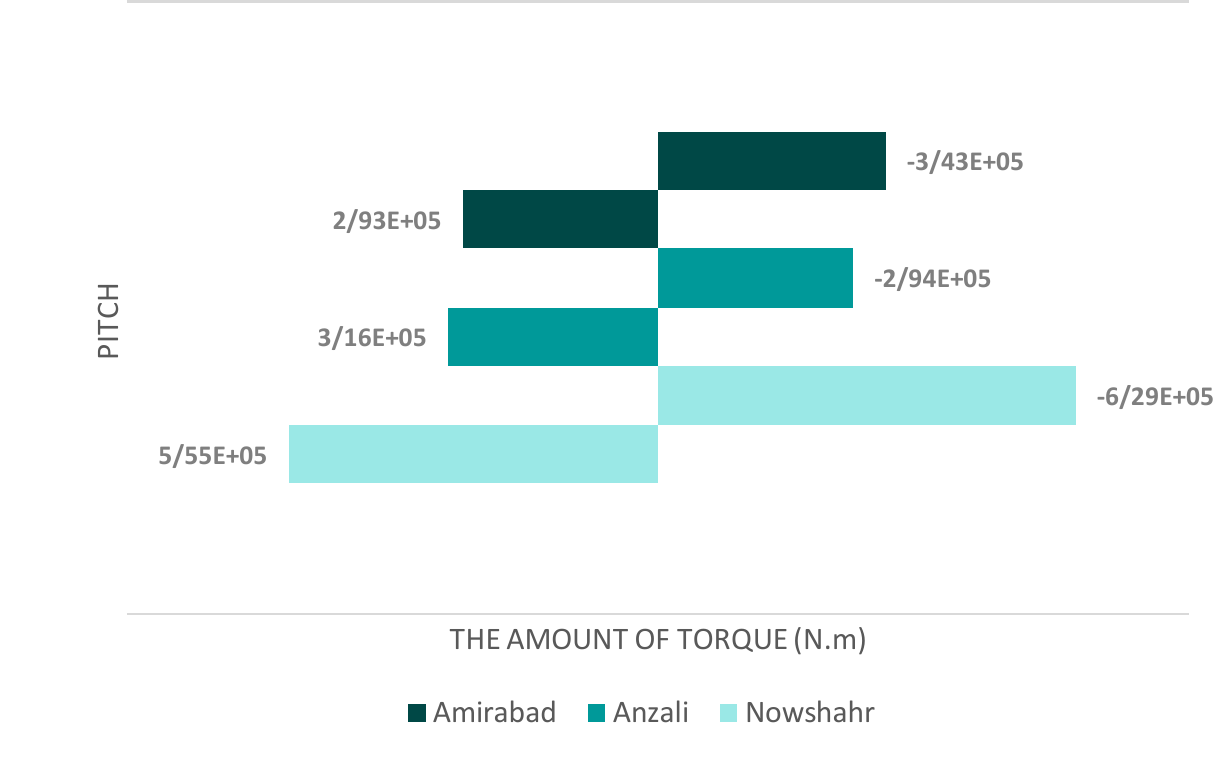}
 	  	}
 	  	\caption{Comparison of the wave torque and power of added mass force both in surge and pitch in both directions. (from up to down: Amirabad, Anzali, and Nowshahr)}
	\label{4_18}
\end{figure}

 It can also be seen that the radiation force of the Nowshahr port converter is much greater than the other two ports, which shows a logical trend in the outputs due to the larger oscillations of the Nowshahr port and its higher oscillation speed. It should note that the maximum wave power of radiation damping force in surge and torque is around 1.42 $MN$ and 1.29 $MN.m$, respectively. Accordingly, it can be said that comparing the radiation force caused by the movement of the converter due to the collision of waves can be considered as a relatively suitable criterion for comparing the performance of the converter.
% \subsection{Added Mass Force Outputs}
In the study of the added mass forces, the two modes of movement of the converter flap backward (collision of a wave crest) and moving the converter flap forward (collision of a wave through) of this force will have different values.
For each port in two degrees of freedom of Pitch and Surge, software outputs are obtained, and the values are presented in Figure \ref{4_18}.

From the outputs of WEC-Sim software, it can be seen that the amount of force and moment of added mass in the Nowshahr port affects the converter flap more than the two other ports. The maximum wave torque and added mass force are approximately 5.55 $MN.m$ and 1.1 $MN$, respectively. Furthermore, the increase in added mass force in the Amirabad port compared to the Anzali port is due to the larger area of the converter flap in this port.
\\
%in ghesmate paiin bashe ya hazf she? ref dade beshe be payanname?
Since the proposed optimal design for Anzali port is 18 meters by 8.7 meters and the Amirabad port is 23 meters by 7 meters, the force and moment of added mass (in a resistant form) in Amirabad port has been achieved more than Anzali port, which indicates that the presence of added mass or moment force on the converter flap is not directly related to the converter's capture factor and should be considered along with other parameters to evaluate the performance of the converter. Besides, the presence of a more amount of force or moment of added mass on a converter flap can not be considered a reduction in its performance.

\begin{figure}[htb]
    \centering
   \includegraphics[height = 5cm, width = 7.5cm]{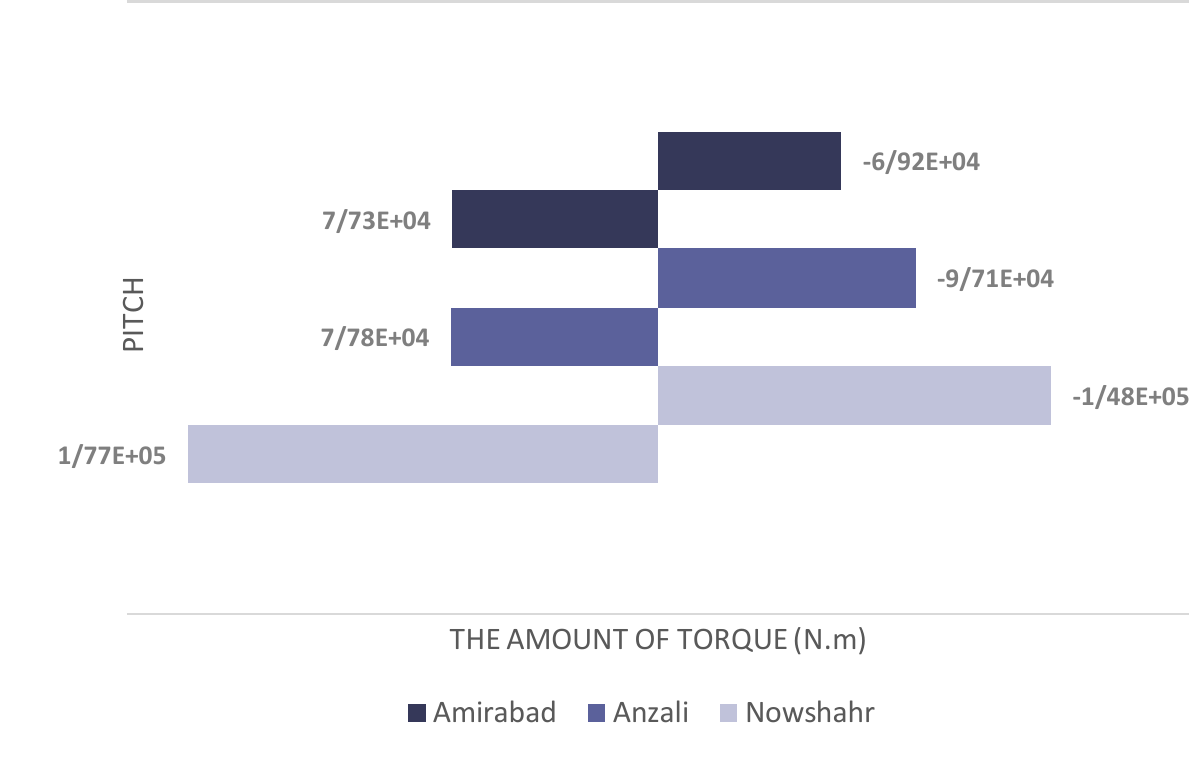}
    \caption{Comparison of the wave torque of PTO force in all locations. (from up to down: Amirabad, Anzali, and Nowshahr)}
    \label{4_19}
\end{figure}

% \subsection{The Outputs of Obtained Torque From Power take-off System }
The two modes of movement of the converter's flap have been shown to evaluate the obtained torque outputs from the power take-off system. Since the power take-off system is practically activated in an oscillating motion, for each port in the pitch degree of freedom, the software outputs are obtained, and the values are presented in Figure\ref{4_19}. 

As shown in the software outputs, the torque amount applied to the energy converter in the Nowshahr port is more than Anzali, and Anzali port is more than Amirabad port. Therefore, it can be expected that this parameter, considering the relationships expressed in the previous chapter, directly affects the efficiency of the converter. Furthermore, according to the transfer rate outputs, it can be expected that more power has been extracted from the Nowshahr port energy converter than the other two ports.

\begin{figure}[htb]
	\centering
	\subfloat [Nowshahr port]{
    	\includegraphics[height = 4cm, width = 12cm]{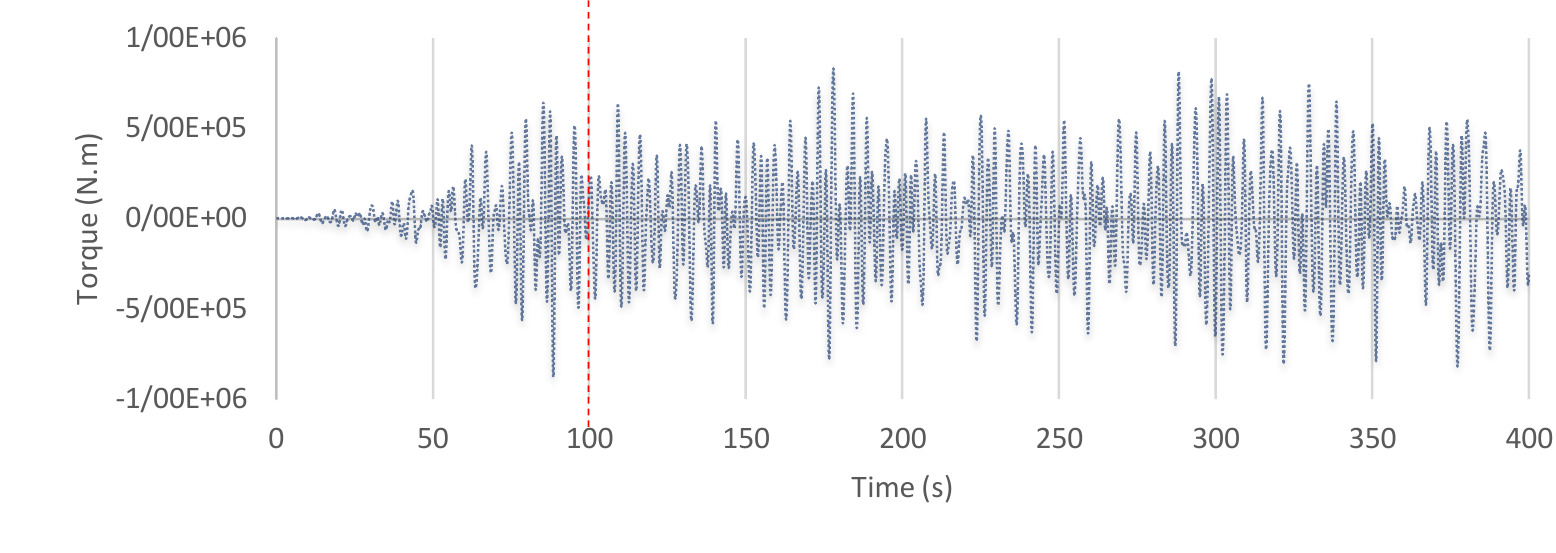}
      		} 
   \vfill
	\subfloat[Anzali port]{
   		\includegraphics[height = 4cm, width = 12cm]{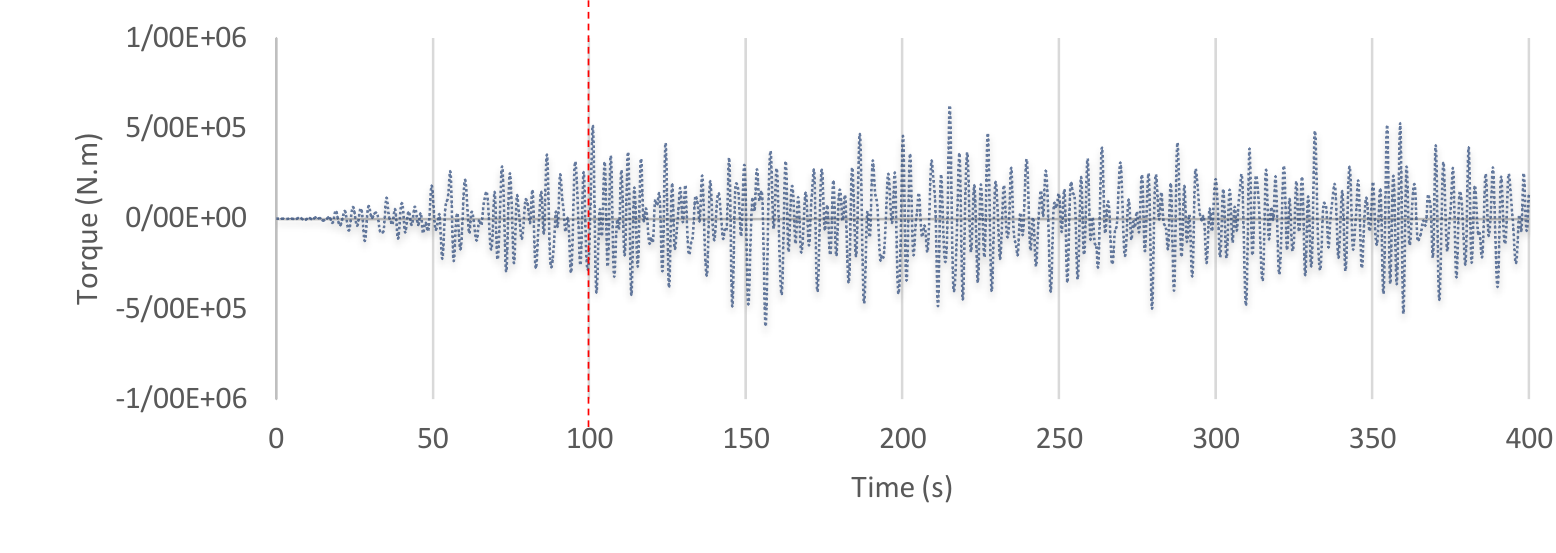}
 	  	}
 	  	\vfill
 	  	\subfloat[Amirabad port]{
   		\includegraphics[height = 4cm, width = 12cm]{4_21.pdf}
 	  	}
 	  	\caption{Total applied torque to converters in studied ports.}
	\label{4_202122}
\end{figure}

Next, the total torque value, which is the sum of all the forces acting on the converter (mentioned in Equation \eqref{26}), is evaluated and compared in the converters considered in the three ports. Figure \ref{4_202122} illustrates the total applied torque values of the converters.

As shown in the torque application diagrams, the torque amount applied to the converter plate in Anzali and Amirabad ports is more than the Nowshahr port. However, it should be noted that merely increasing the amount of torque applied to the converter flap will not increase the energy extracted. The displacement of the converter flap must be considered in conjunction with other environmental conditions. The maximum values of the actuator torque affecting the pressure center of the converter flap (when moving under pressure due to the impact of the wave crest) in Nowshahr, Anzali, and Amirabad ports will be 832 kN, 630 kN, and 621 kN, respectively, in each time step. Also, the maximum values of return torque affecting the pressure center of the transducer plate (during the return movement under reduced pressure due to the wave through impact) in Nowshahr, Anzali, and Amirabad ports are equal to 881 kN, 598 kN, and 588 kN, respectively, in each time step.
\subsection{Comparative Results of Capture Factor}
\begin{figure}[htb]
    \centering
    \includegraphics[width=0.9\linewidth]{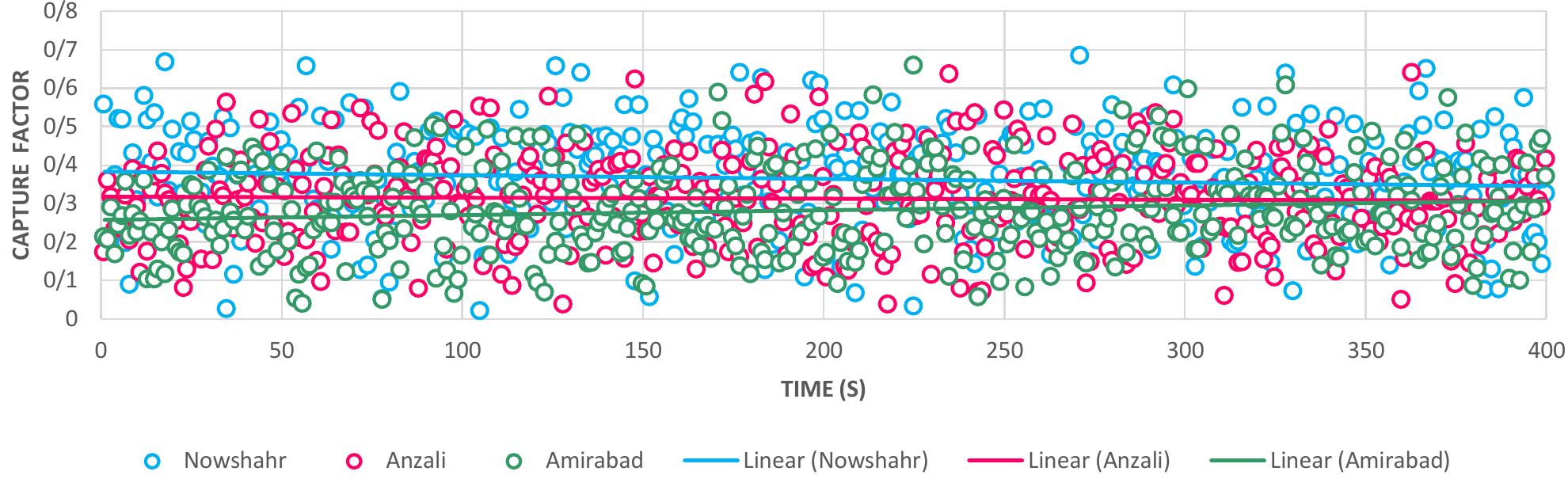}
    \caption{Comparison of capture factor in all locations.}
    \label{4_23}
\end{figure}

According to the application of irregular shortwave theory, the capture factor's value will change at each time step according to the proposed equations. For this purpose, the capture factor's values in all three ports are calculated per time steps in a total run-time of 400 seconds and compared in Figure \ref{4_23}. 
\\

A comparison of the capture factor's values in the form of the above diagram indicates that this quantity in Nowshahr port is more than the other two ports. It is worth mentioning that all these values have been calculated at points of each port with the maximum impact wave energy. The exact values of the capture factor in all three ports, the total energy values of the waves, and the energy absorbed by the converter in each of the ports are examined in the table \ref{tab:Table 4-4}.

\begin{table}[htb]
    \small
\begin{center}
\begin{tabular}{cccc}
  \toprule
\textbf{Port's Name} & \textbf{Capture Factor (\%)} & \textbf{Incident wave Energy ($Kw$)} & \textbf{Absorbed Energy ($Kw$)} 
\\ \hline	
 Nowshahr &  62.99 & 5376.66 & 3358.76  \\ 
\hline
Anzali  & 48.84 & 4471.66 & 2183.96  \\ \hline
 Amirabad & 42.17 & 4116.66  & 1735.98 \\ \bottomrule
	\end{tabular}
 \caption{The results of absorbed power, capture factor and incident wave power in studied sites.}
    \label{tab:Table 4-4}
    \end{center}
\end{table}

It can be said that in the studied ports, Nowshahr port has the priority of constructing a wave energy converter system, considering the comparison of the capture factor values. Then, Anzali and Amirabad's feasibility on the southern shores of the Caspian Sea can be studied.

Simultaneous consideration of converter design parameters, including environmental and geometric quantities, can help achieve maximum system power and significant progress in using this type of energy in Iran's marine industry in the future. Concerning the WEC-Sim module's suggestion to take 1, 2, 1.8, and 2.3 meters of transverse distance between elements in the $Y$ axis in Nowshaher, Anzali, and Amirabad ports, respectively.

\subsection{Flap Response}

\begin{figure}[htb]
	\centering
	\subfloat [Nowshahr port]{
    	\includegraphics[height = 4cm, width = 12cm]{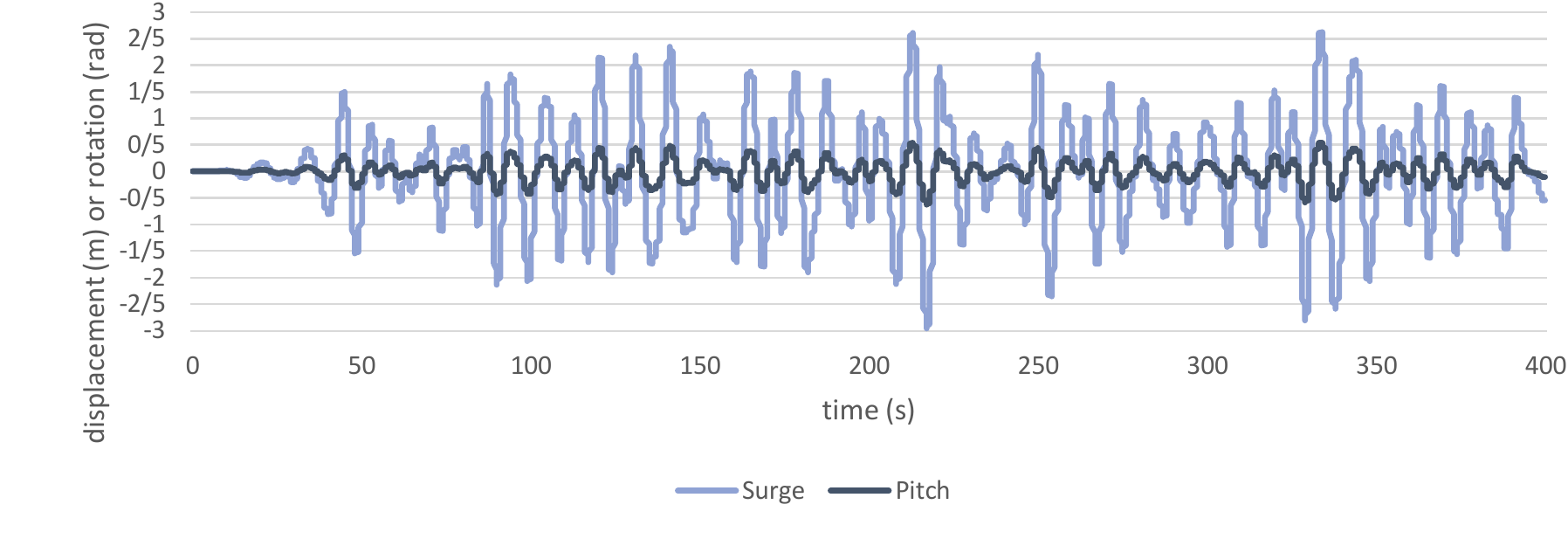}
      		} 
   \vfill
	\subfloat[Anzali port]{
   		\includegraphics[height = 4cm, width = 12cm]{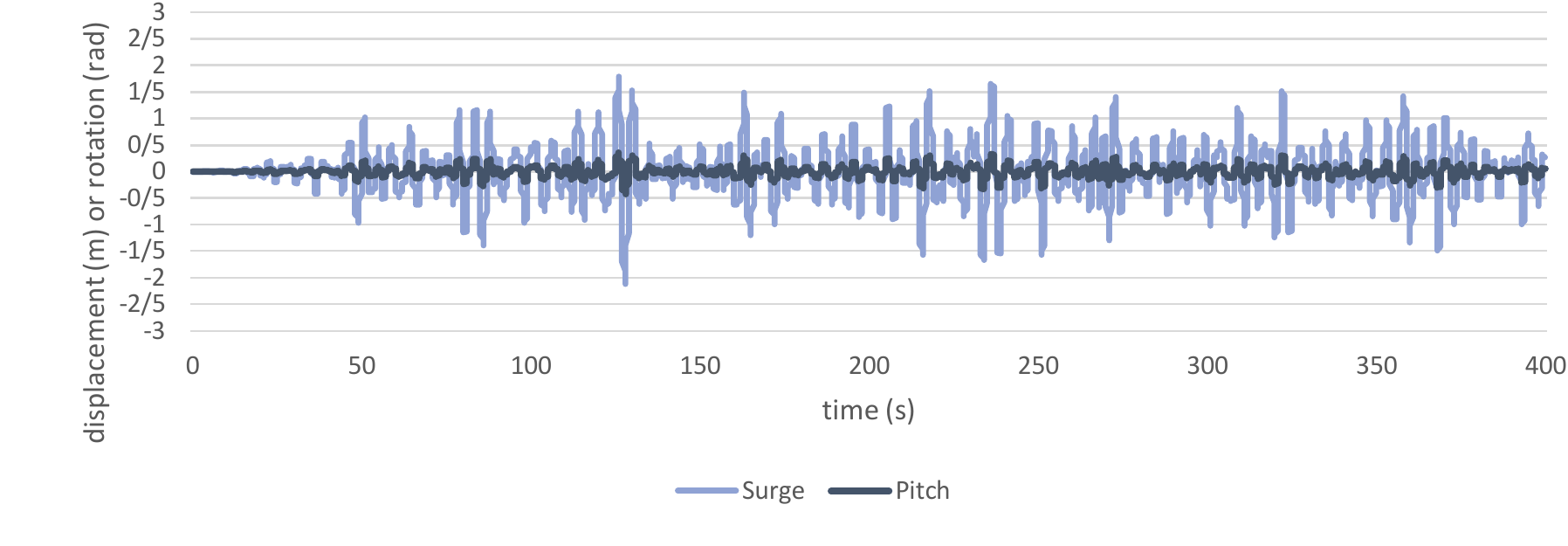}
 	  	}
 	  	\vfill
 	  	\subfloat[Amirabad port]{
   		\includegraphics[height = 4cm, width = 12cm]{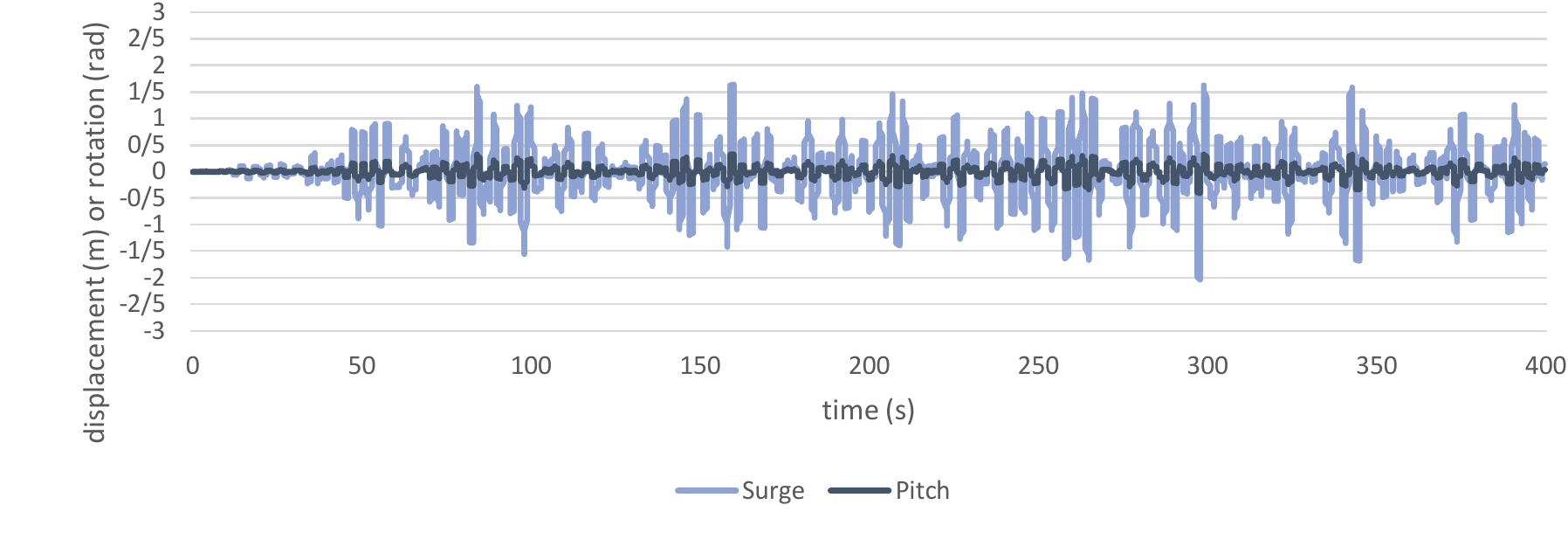}
 	  	}
 	  	\caption{The amount of displacement and rotation of the flap in all locations}
	\label{4-131415}
\end{figure}

In this section, the amount of flap displacement or rotation responses to the implied is investigated in Pitch and Surge degrees of freedom. It must be noted that the total run-time is 400 second and the ramp time is 100 second, and the time step is chosen 0.1 regarding the Courant number \cite{15}. The amount of fluctuation, which can be seen in figure \ref{4-131415} for Surge and Pitch, is given in meter and radian, respectively.

It is clear that oscillation in Nowshahr is by far higher than the other sites, and this difference is also witnessed in the displacement of the flap and the angle of rotation. The maximum force of the flap in the Nowshahr port (during the collision of wave crest) is 2.624 meters with a 0.54-radiant maximum rotation. These measures for Anzali and Amirabad ports are 1.78, 1.64 meters, and 0.34, 0.32 radiant, respectively. Furthermore, when the flap moves because of wave trough, the maximum amount of the flap's displacement is 2.96, 2.12, and 2.04 meters. The maximum angle of rotation is 0.619, 0.42, and 0.4 radiant in Nowshahr, Anzali, and Amirabad. Finally, the flap's rotational velocity is 0.025, 0.02, 0.021 radian per second in Nowshahr, Anzali, and Amirabad, accordingly. 

\section{Conclusions}
\label{Sec-conclusions}
Firstly, the oscillator surge converter chose to expand the horizon of development, installing and maintenance cost, converter movement ability, and the converter industry's popularity. Next, designing geometrical parameters of the converter depend on environmental(sea-state) conditions. Thus, the research locations are chosen carefully, then their parameters are assessed for implementing a model to extract the most potential energy. Thirdly, the WEC-Sim module seems to be a better choice than its rivals because it takes the standard formats of geometrical and environmental files. It is an open-source and user-friendly module in MATLAB software where itself is one of the popular ones for implementing codes. According to the literature review, Nowshahr, Anzali, and Amirabad are chosen to install an oscillating wave surge converter(OWSC) in the Caspian sea. Due to the broad set of data in the Caspian Sea, an ocean metadata searching algorithm is written to find desired data. The height of the designed converter's flap is from 7 to 8.7 meters, and the width is between 18 and 23 meters. The converter's fluctuation and applied torque show that the flap's total used torque is high when more wave height is observed in the installation site. It can be said that total power encountering the oscillator's flap, together with the capture factor, can be considered a determinative parameter for revealing the system's efficiency. Finally, the simulation results indicate that the Nowshahr port parameters have the maximum potential for harnessing energy by oscillator wave surge converter with an efficiency of 63 percent and with the highest absorption coefficient.

\section*{Acknowledgement}
The authors would like to express their gratitude to the Iranian National Institute for Oceanography and Atmospheric Science (INIO), which provided the data needed for performing this project. Also, we shall more specifically appreciate the support of Dr. Mehdi Neshat, who was of great help, valuable comments, and useful suggestions in this research. This work has been supported by the Iran Port and Maritime Organization under the grant number TE178.

\end{document}